# Dark Matter - Matter That Can't Be Touched

Team Astrophysics

**Anukritee Negi**[1], **Shivani AC**[2], **Srijan Rauniyar**[3], **Krishna. S. Kamath**[4], **Richeek Debnath**[5]

[1] 10th Grade, St.Joseph's Convent School, Kotdwara, Uttarakhand,India

[2] Grade 11 (ISC), K'sirs International School, Coimbatore, Tamil Nadu, India

[3] High School 2nd Year Student, Arniko College, Biratnagar,Nepal

[4] High School Student,Space Enthusiastic

[5] Second Year Undergraduate Student ,Department Of Chemical Sciences,IISER Berhampur,Transit Campus Govt(ITI) Building,Engg School, Berhampur,Odisha

## Abstract

During the last century many observations have been made to peep into the DARK MATTER in the universe and many astonishing behaviours of Galaxy clusters have been found which do not fit to any theories formulated before. However, Optical Spectroscopic observation has been initially used to measure the rotational velocity of the Andromeda Galaxy as a function of distance found in contrast to the Law of Gravitation. Another observation of X-ray of luminous gas of an Elliptical Galaxy has been carried out where bending of light emitted from a cluster's gravitational field was studied. Since the 1950s, the Big Bang cosmology scenario has held the leading position as the most successful model for the origin and evolution of the Universe. Due to the expansion of space, electromagnetic radiation emitted in the distant Universe is redshifted on its path towards us which can be used to determine the distances of astronomical objects.

The lingering radiation from the Big Bang that permeates the whole Universe is the CMB or cosmic microwave background, which can be utilised to 'travel back in time the farthest'. The cosmologists tried to find out nature and the content of the Dark Matter which was found to be 98.3% of the Low speed mass and only 1.67%. Moreover, most of the Dark Matter contribution came out from Non baryonic nature.

Different models have been proposed to see its nature and its time evolution viz Cold, Warm, Mixed Dark Matter Models, Self Interacting and Self Annihilating, Fuzzy and Modified Gravity models. Its constituent particles include few Baryonic and some Non-Baryonic particles.

And finally detections were made, though no direct evidence are available but strong indirect evidence of its presence have been confirmed. This paper is an attempt to look into the aforesaid perspectives of Dark Matter and make predictions to connect the observed properties of galaxies.

---

## 1. Introduction

From the last century Astronomers have been observing a major portion of the universe which has mass but isn't emitting any light or it is very faint. It is called DARK MATTER. Its first indirect detection was given by Zwicky(1933) in his studies on “Coma Cluster” where velocity dispersion of the Luminous matter was found to exceed ~50 times for having that amount of mass.

His work was followed up by Smith on the Virgo cluster of Galaxies where he found a high mass to light ratio. But there were some other arguments for these types of unusual behaviours; taking into consideration , the Giant Ionised Gas clouds which emit faint light. [1]

But the Optical Spectroscopic observation by Babock(1939) put a fullstop on this argument. He measured the rotational velocity of the Andromeda Galaxy as a function of distance and found large velocities even at large radius in contrast to the Law of Gravitation. But the Light Intensity falls off naturally.

Apart from these observations many others were also observed: Observation of X-ray luminous gas of an elliptical galaxy, bending of light emitted from a cluster by gravitational field (Yes, gravity bends light!!). [8]

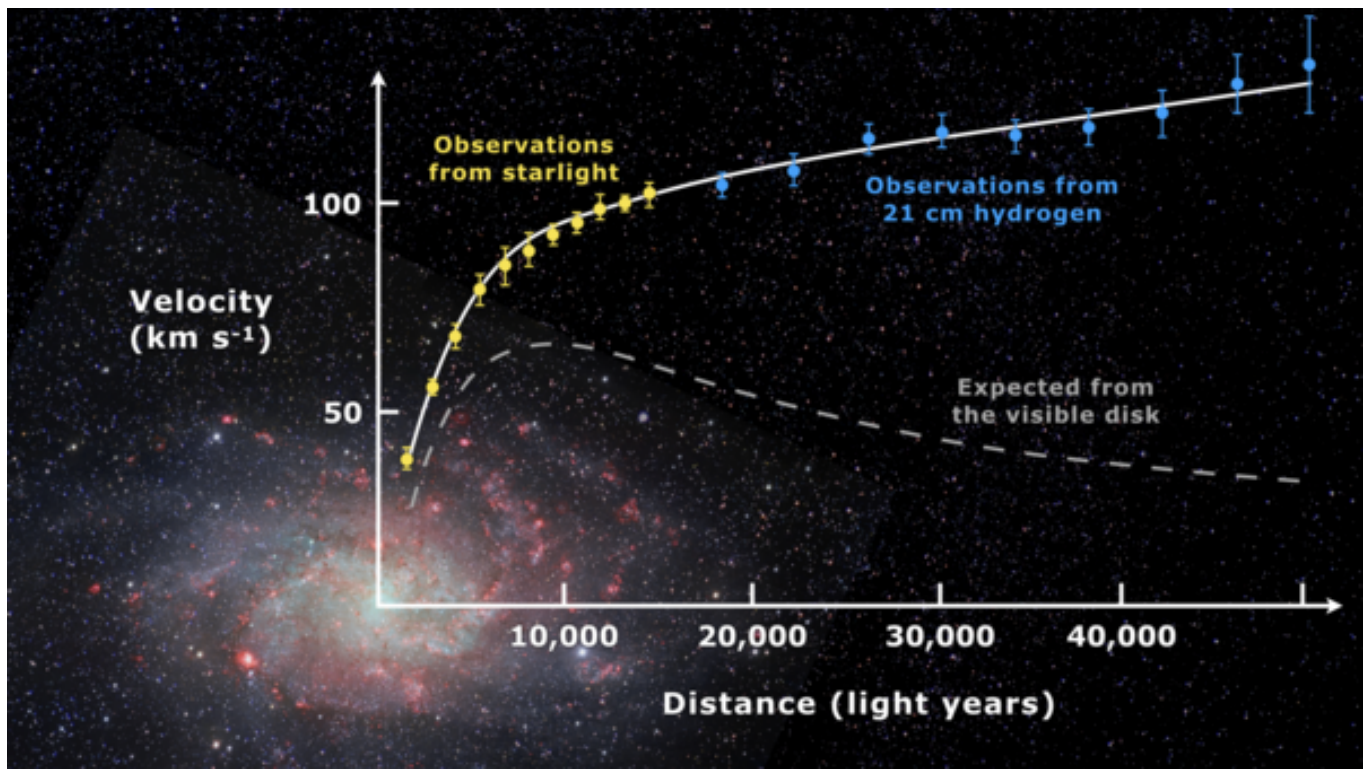

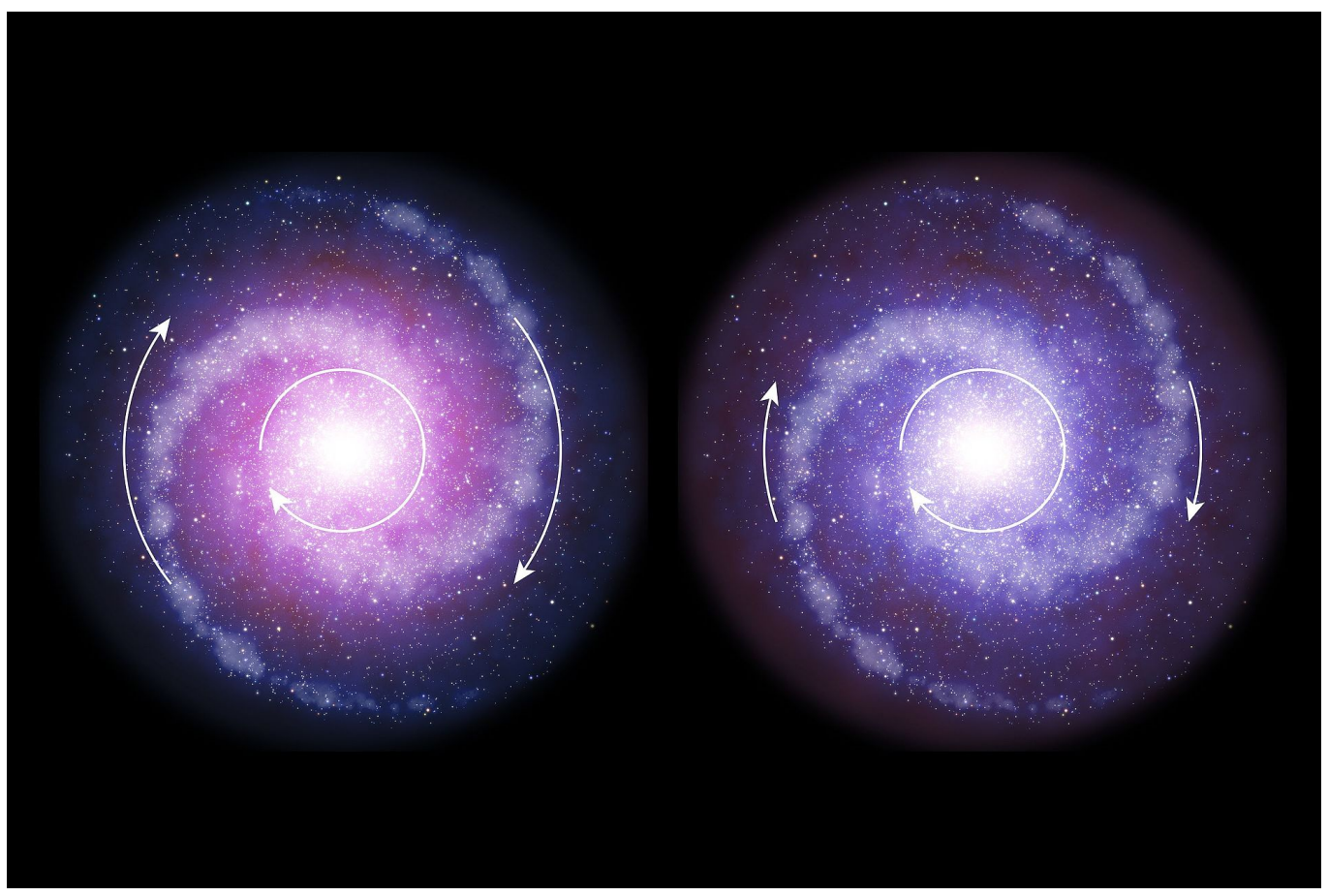

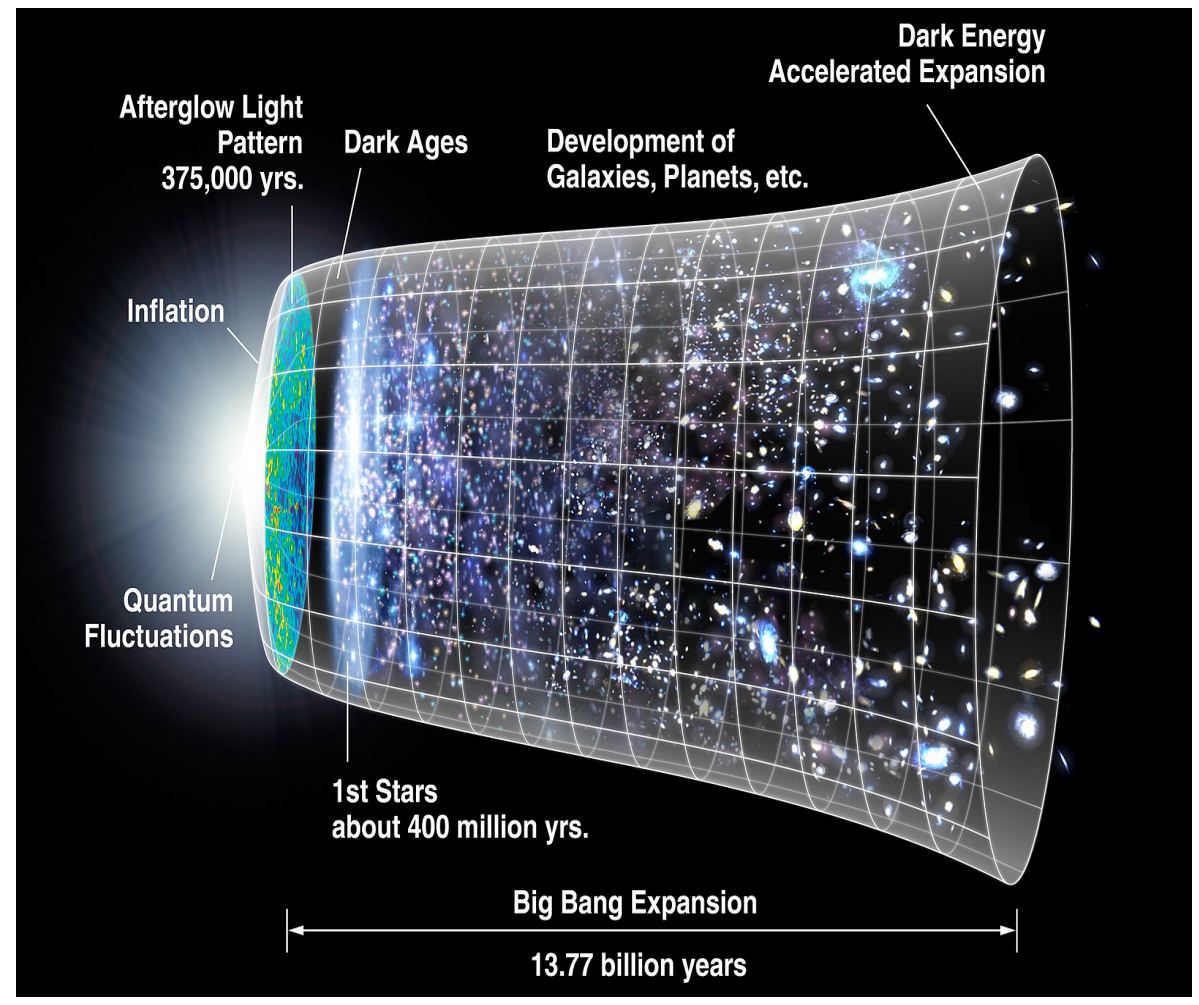

## 2. Big Bang Cosmology

Since the 1950s, the Big Bang scenario has held the leading position as the most successful model for the origin and evolution of the Universe. In this cosmology, the Universe started out extremely hot and dense some 14.1+1.0 −0.9 Gyr ago (Tegmark et al. 2004). Early on, there were no galaxies, no stars and no planets. The Universe was instead filled by a gas of subatomic particles at an extremely high density. As space expanded, the energy density dropped and the cosmic plasma cooled. After about 3 minutes, the Universe had cooled sufficiently to allow synthesis of the light elements H, He, Li and Be.

Due to the expansion of space, electromagnetic radiation emitted in the distant Universe is redshifted on its path towards us, so that the observed wavelength of light, $\lambda_{obs}$, is larger than the wavelength at which it was emitted, $\lambda_{emit}$. The longer the light path through the expanding cosmos, the larger is the amount of redshift induced. Redshift can therefore be used to determine the distances to astronomical objects [9,10]. The redshift, $z$, is defined to be:

$$z = \frac{\lambda_{obs}}{\lambda_{emit}} - 1 = \frac{a_{obs}}{a_{emit}} - 1$$

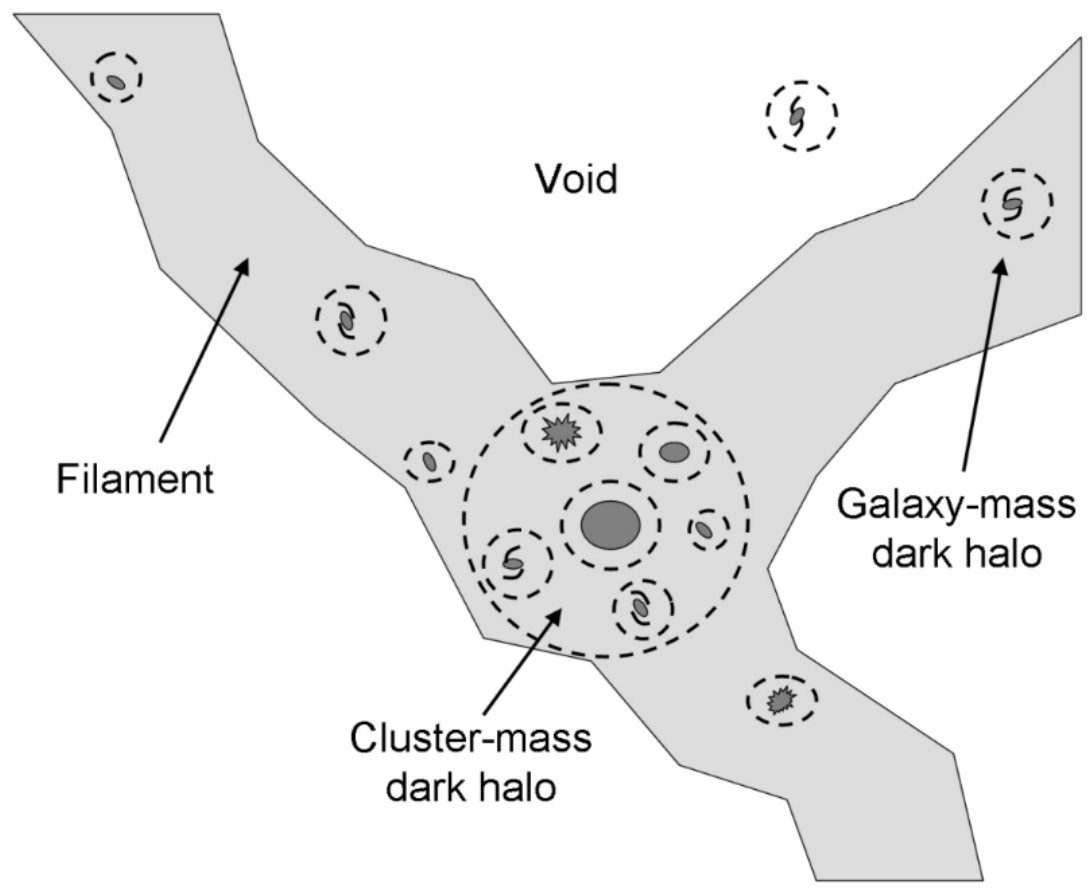

## 3. Cosmic Microwave Background

**What is the Cosmic Microwave Background?**

The CMB or cosmic microwave background, is the lingering radiation from the Big Bang that permeates the whole Universe. Nearly 14 billion years ago, when the Universe first began, it was a hot plasma of particles, primarily protons, neutrons and electrons, and photons (light). Particularly, the photons were continually interacting with free electrons during the first 380,000 years which prevented them from travelling far. That indicates the Early Universe was dim, similar to fog. [11,12]

**Why is it so important to study?**

We can travel back in time the farthest utilising the light in the cosmic microwave background. It emerged roughly 380,000 years after the Big Bang, and the seeds from which the stars and galaxies we can now see eventually developed are visible on it. A complicated narrative that explains the History of the Universe both before and after the CMB was released is concealed in the radiation pattern. [11,12]

**When was it first detected?**

They estimated that the temperature of this radiation is only a few degrees above absolute zero (5K), which is the wavelength of a microwave, as a result of the Universe's expansion. It wasn't until 1964 that it was accidentally discovered by Arno Penzias and Robert Wilson in New Jersey while utilising a sizable radio antenna. For this discovery, they were given the Nobel Prize in Physics in 1978. [11,12]

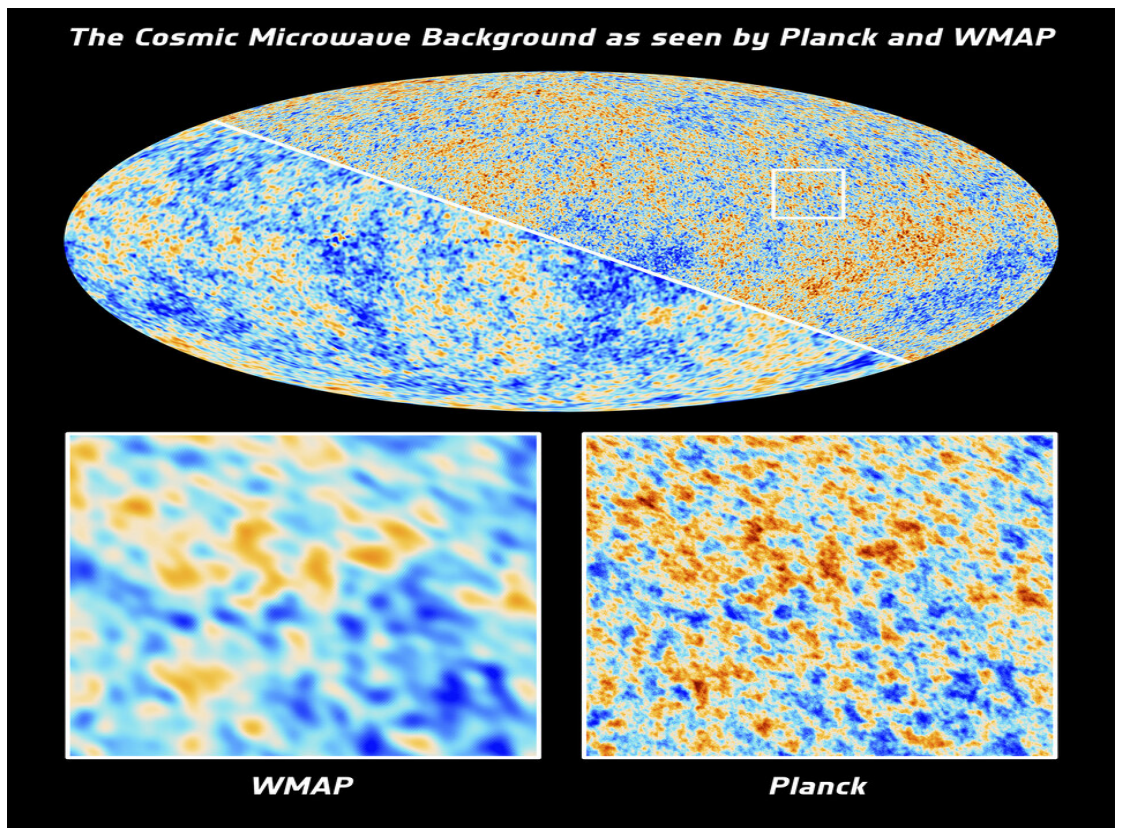

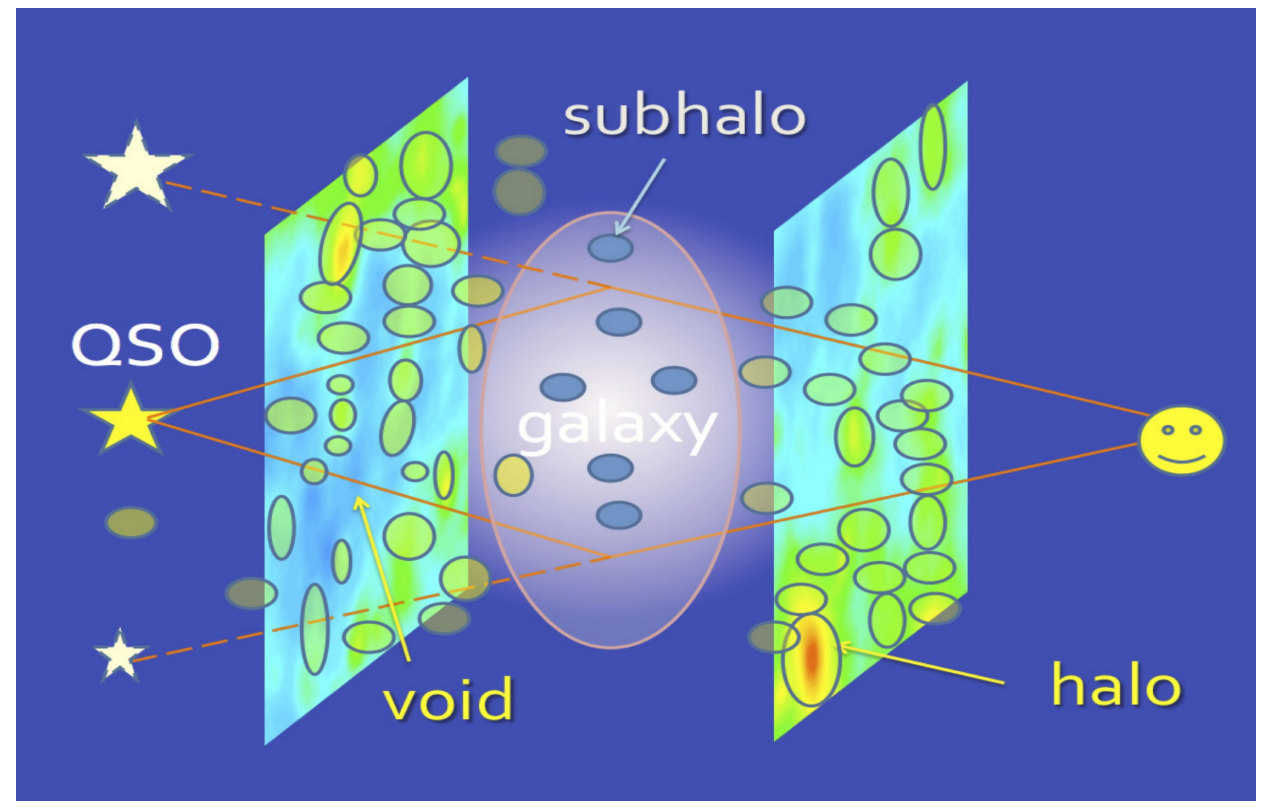

## 4. Nature Of Dark Matter

When Dark Matter's presence was confirmed then cosmologists tried to find out its nature. They studied the density profile of the Universe and got:

$$\rho(total) = \rho(Low\ speed\ mass) + \rho(High\ speed\ mass) + \rho(\lambda)$$

where ρ is density and λ is just a constant. But the high speed mass contribution is very negligible. So,

$$\rho(total) \approx \rho(Low\ speed\ mass) + \rho(\lambda).$$

Further investigations on Low speed mass showed that it indeed had the Dark Matter contribution and it contains both the Baryons and Non- Baryons.

$$\rho(Low\ speed\ mass) = \rho(Baryonic) + \rho(Non\ Baryonic)$$

So, this implied,

$$\rho(Low\ speed\ mass) = \rho(Normal\ Matter) + \rho(Dark\ Matter)$$

and the content of the Dark Matter was found to be 98.3% of the Low speed mass and only 1.67%. Moreover most of the Dark Matter contribution came out from Non baryonic nature.

Thus, the nature of the Dark Matter was found to be Non-Baryonic though there is always some Baryonic contribution.

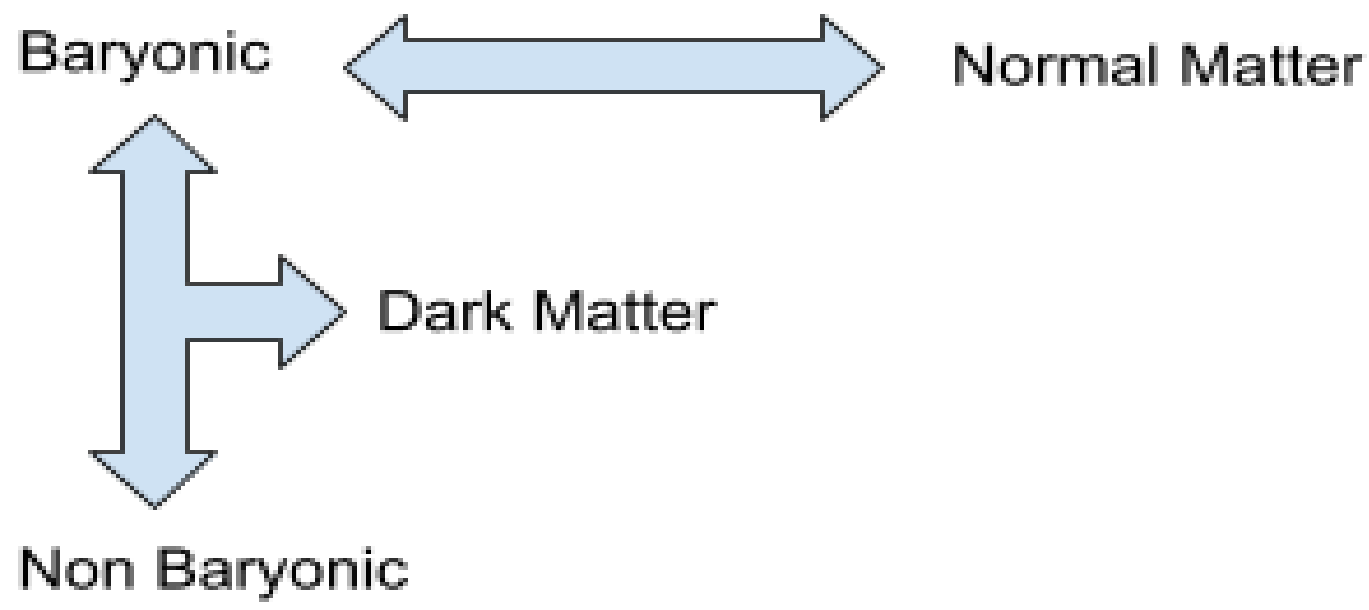

## 5. Different Models

After the nature of the Dark Matter was confirmed then different models were proposed. The most likely candidates are listed below.

**I. Cold Dark Matter Model(CDM):**

This model refers to the Dark Matter moving at low speed (i.e $v <<<< c$ ). The model also lies with the observations : It correctly predicts the down-up scheme of structure formation of the Universe i.e formation of small mass stars followed by the formation of the larger structure. Moreover CDM particles are assumed to be collisionless, long lived, and behave as a perfect fluid on a large scale.

**II. Warm Dark Matter Model:**

This model put the Dark Matter's speed in between Low speed mass and Very High speed Mass(called Hot Dark Matter). Basically it resolves one of the main problems in CDM which couldn't predict the central density of Dark Matter Halos correctly.

**III. Mixed Dark Matter Model:**

This refers to the mix state of both Cold Dark Matter and Hot Dark Matter. But after introduction of $\lambda CDM$. This model remains of less concern.

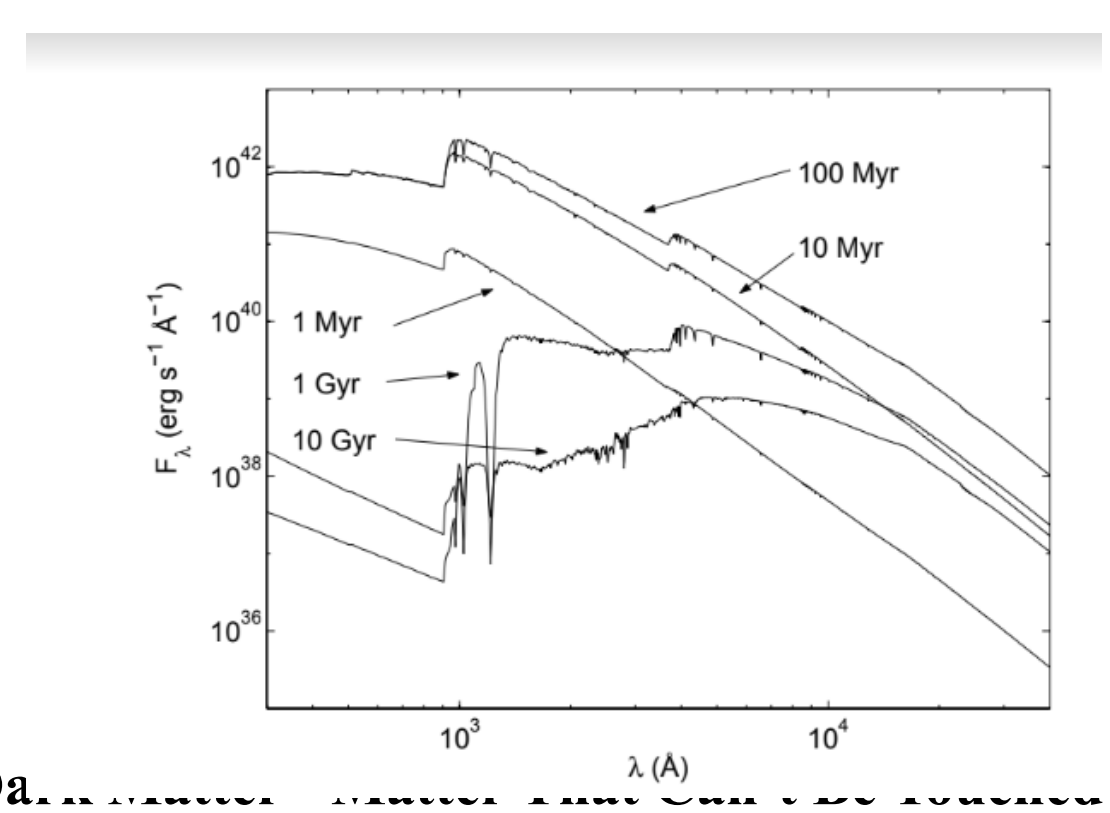

*The spectral evolution predicted by Zackrisson*

**IV. Self Interacting Dark Matter:**

This model is based on non-gravitational interaction of Dark Matter particles.

## V. Self Annihilating Dark Matter:

It is based on the annihilating nature of Dark Matter when two Dark Matter particles collide.

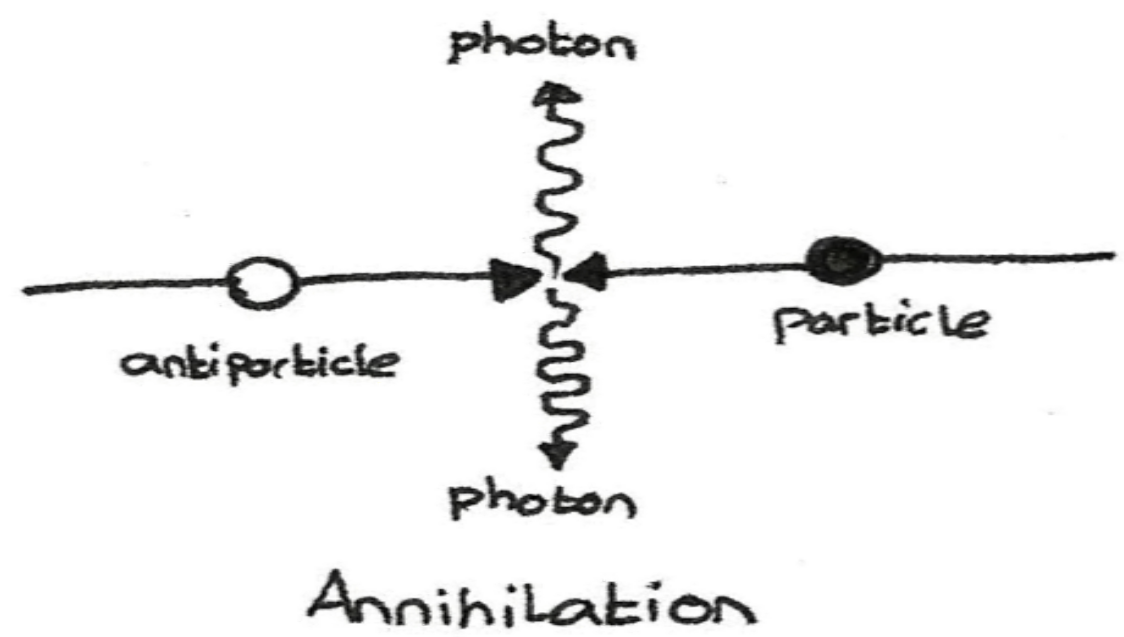

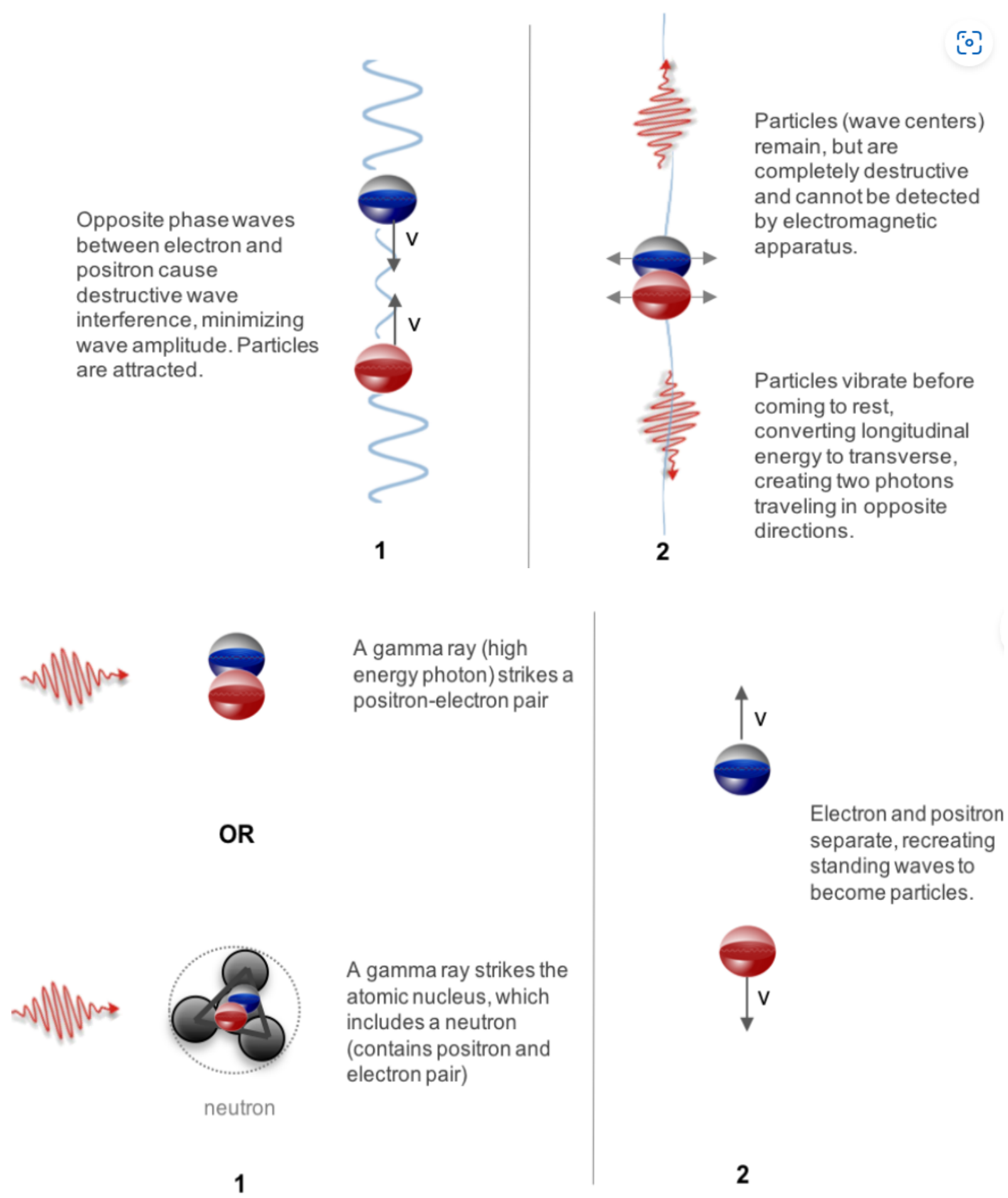

*Schematic representation of annihilating nature of dark matter*

Adding to that, this part solves the Central Density problem of a Dark Matter Halo. After annihilating they form radiation or high speed particles(relativistic) which upon detection can help to know about the deep inside the universe. One of the similar detection is done by AMS-02 which has captured flux of antiprotons. It is expected to come from Dark Particles' collision.

**VI. Fuzzy Dark Matter Model**

This states Dark Matter particles with small masses but effectively large size.

**VII. Modified Gravity Model**

Apart from the above mentioned ones, there is another model which doesn't even consider the presence of Dark Matter called the Modified Gravity Model. It includes a new type of dynamics called Modified Newtonian Dynamics (MOND). [13]

---

## 6. Dark Matter Candidates

Now, the question arises: What is this Dark Matter Particle?

Well, there are many candidates for Dark Matter particles. Some of them are Baryonic and some of them are Non-Baryonic. Baryonic candidates include:

1. Faint stars and stellar remnants
2. Cold gas clouds
3. Warm/Hot intergalactic gas
4. Rydberg matter

And Non-Baryonic ones include

1. Neutrinos
2. Axions
3. Supersymmetric Particle
4. Mirror Matter
5. Primordial Black Holes
6. Preon Stars
7. Quark Nuggets
8. WIMPzillas
9. Matter in parallel branes
10. Dark energy as dark matter

Apart from these two categories there are 2 more strong candidates;

1. WIMPs(Weakly Interacting Massive Particles)
    a. Small, Non- baryonic, but having high range mass.
    b. Interact through Gravity and Weak Nuclear Force.
    c. Lack of strong interaction with normal matter and with electromagnetism.

2. MACHOs(Massive Astrophysical Compact Halo Objects)
    a. Heavier than WIMPs, it can constitute all the Dark Matter.
    b. Both Non-Baryonic and Baryonic can behave as MACHOs.

---

## 7. Detection Of Dark Matter

Although the nature of Dark Matter is difficult to find, some indirect detections were claimed to be made:

a. It has been detected based on the WIMP flux as the Earth moves through this wind.
b. During June the velocity of Earth around Sun is added with velocity of Sun around the Milky Way – giving maximal flux and During December these two velocities act in opposite directions giving minimal flux. This was detected by the DAMA observatory in an underground mine at Gran Sasso, Italy.

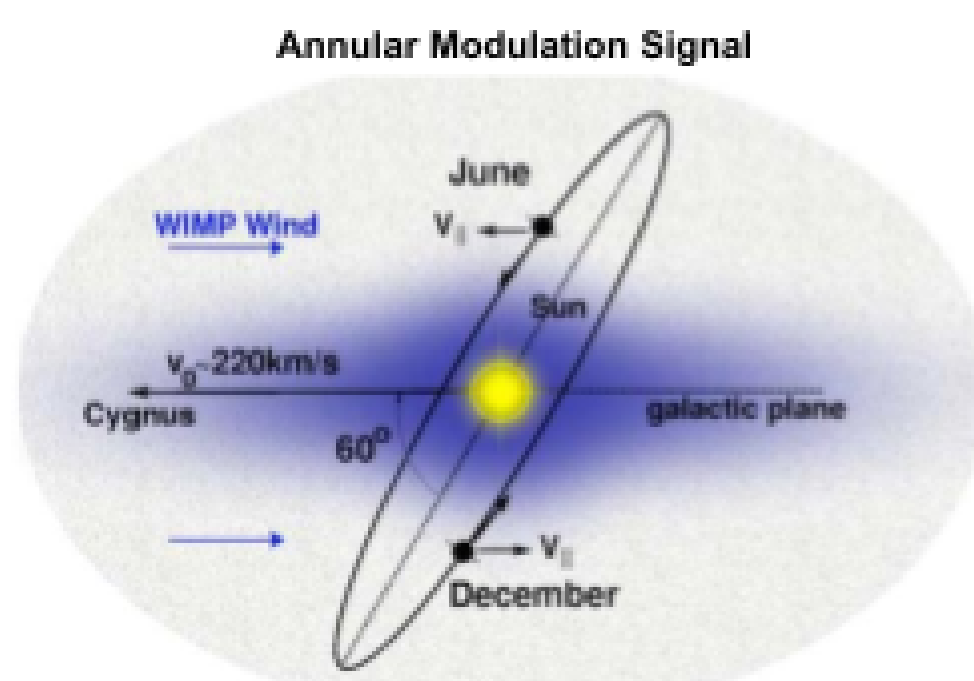

c. Detection by using particle accelerators.
d. Detection by the Absorption of Dark Matter's Energy when Dark Matter Particles interact with target material.
e. Dark Matter particles of mass in high range can be detected by Scattering process.
f. Through the Microlensing event we also have indirect evidence of Dark Matter.

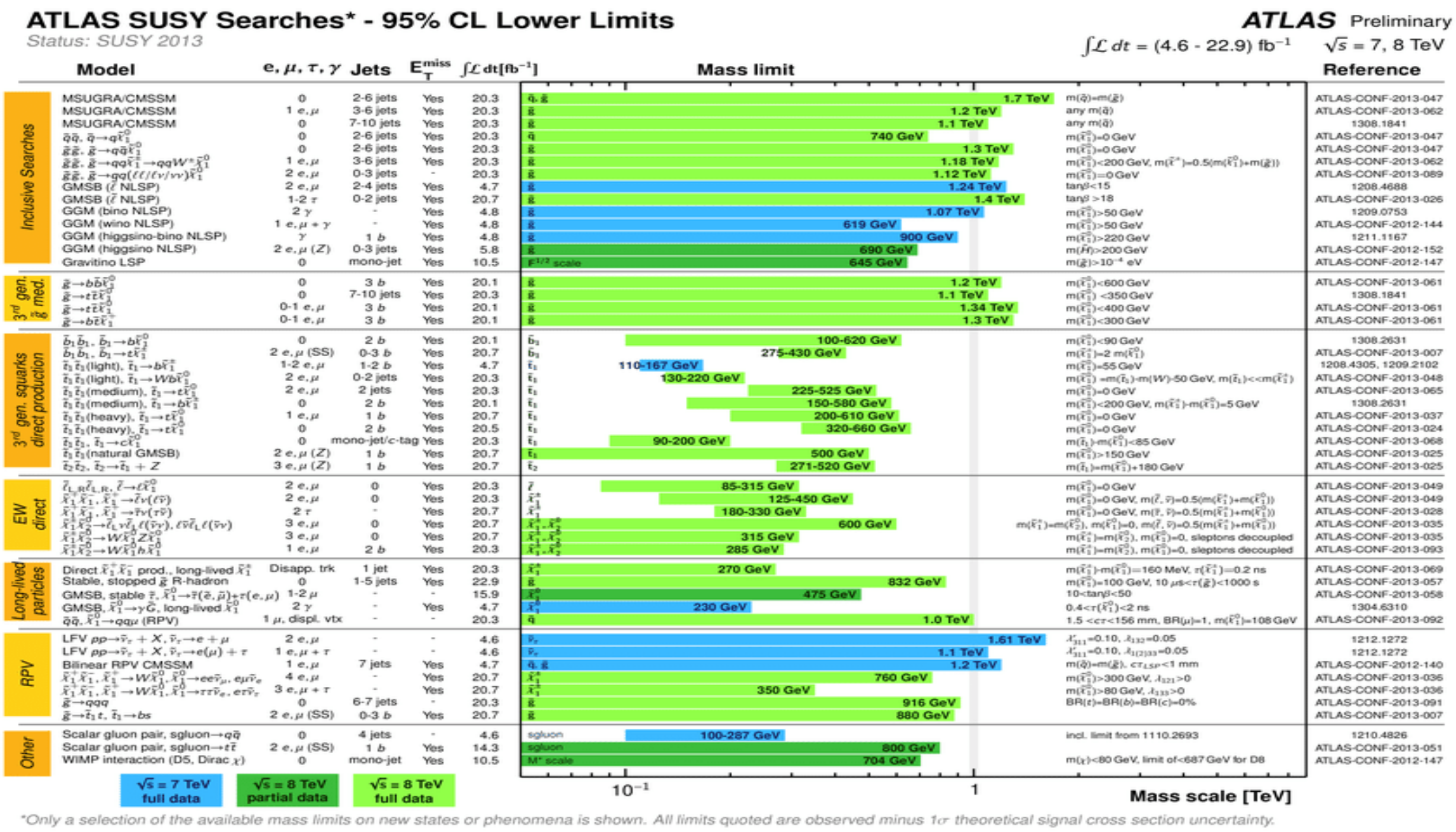

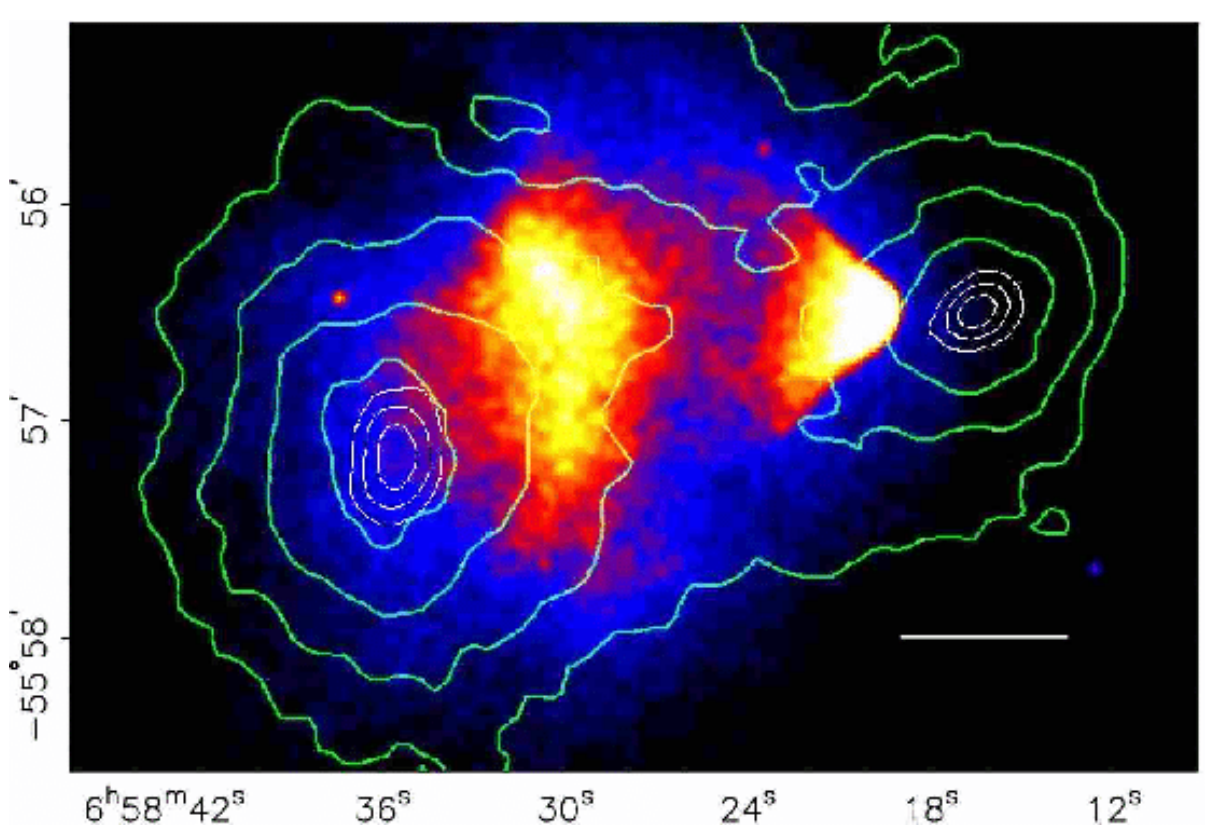

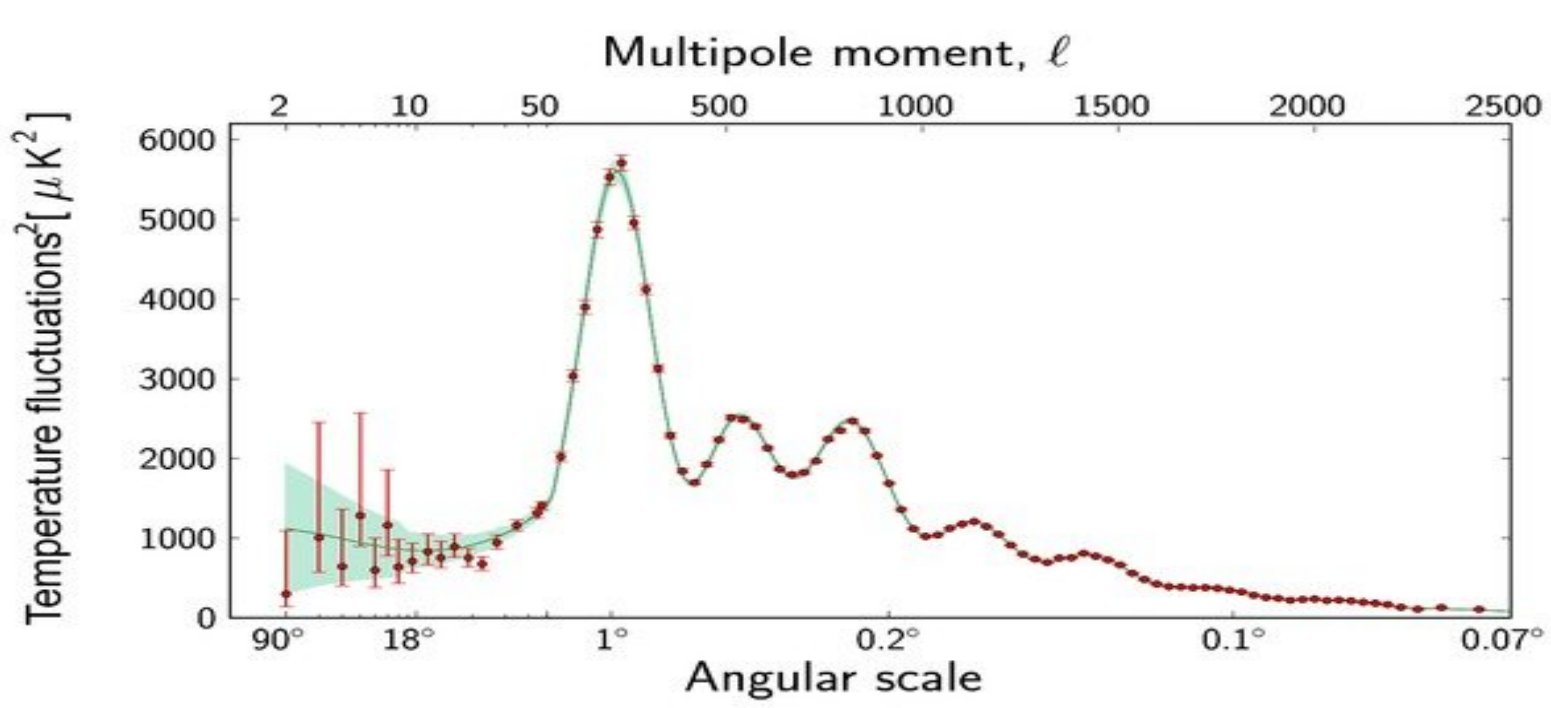

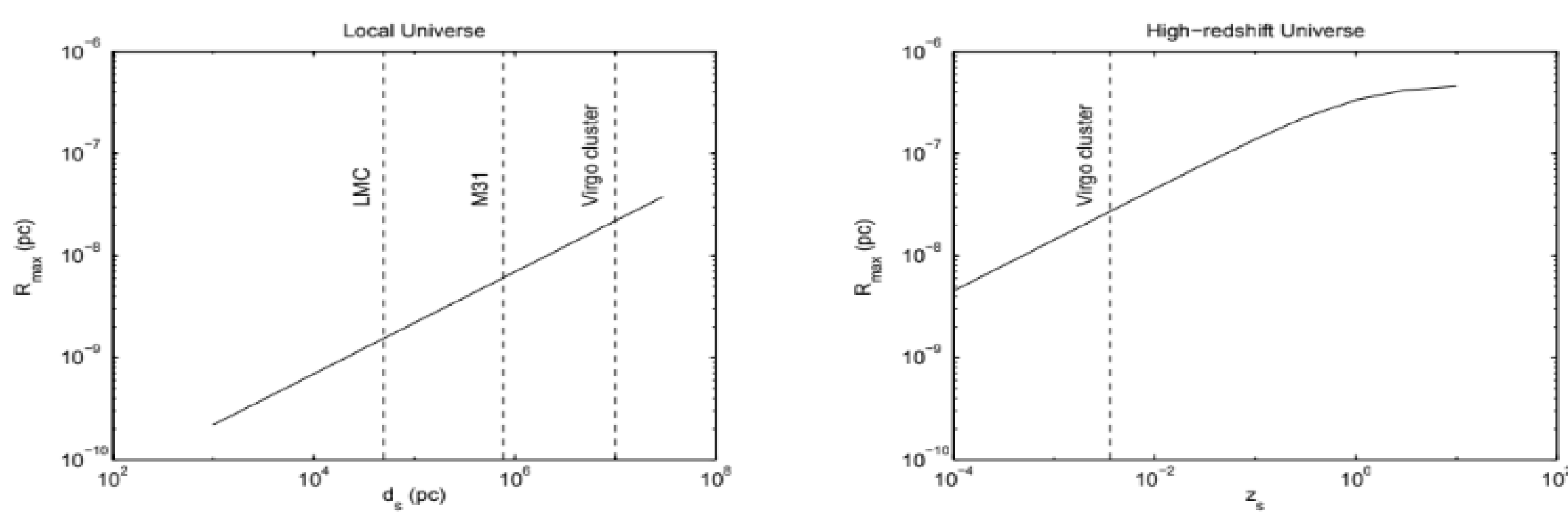

*The maximum radius $R_{max}$ that a compact object of mass M*

## 8. Future Prospects

Although the observational constraints on dark matter halos have substantially improved with the wide-spread use of optical, long-slit rotation curve data as a complement to low-resolution HI rotation curves, this technique is still fraught with uncertainties (e.g. Swaters et al. 2003). Many of the problems associated with non-circular motions can be better addressed by two-dimensional velocity fields obtained through multiple-fibre spectrographs or Fabry-Perot interferometers, and several investigations of this kind have already been instigated (e.g. Blais-Ouellette et al. 2001; Simon et al. 2004). So far, no universal halo profile does however seem to emerge, indicating that the scatter is cosmic rather than due to observational problems.

One of the big future challenges in the game of putting the CDM predictions to the test will be to connect the observed properties of galaxies to those of their dark halos. Do the observationally established halo properties correlate with colour, central surface brightness (disk or actual), baryonic mass, disk scale length, disk thickness or other properties? Until these very complicated issues are resolved, the bias introduced by testing the average CDM halo properties against the dark halos of some particular class of objects – such as dwarf galaxies or LSBGs – will be hard to control. Stronger constraints on the influence of dark baryons on the observed kinematics of galaxies are also badly needed.

On the theoretical side, the resolution of N-body CDM simulations needs to be increased to the level of current observations, so that the severity of the central core/cusp discrepancy in galactic halos becomes clear. The simulations furthermore need to evolve from the stage of providing predictions for the spherically averaged, dark halo density profiles to giving predictions for the kinematics of baryons inside the actual (triaxial) potentials of these dark halos. Although certain steps in this direction have already been taken (e.g. Hayashi et al. 2004), there is still a long way to go.

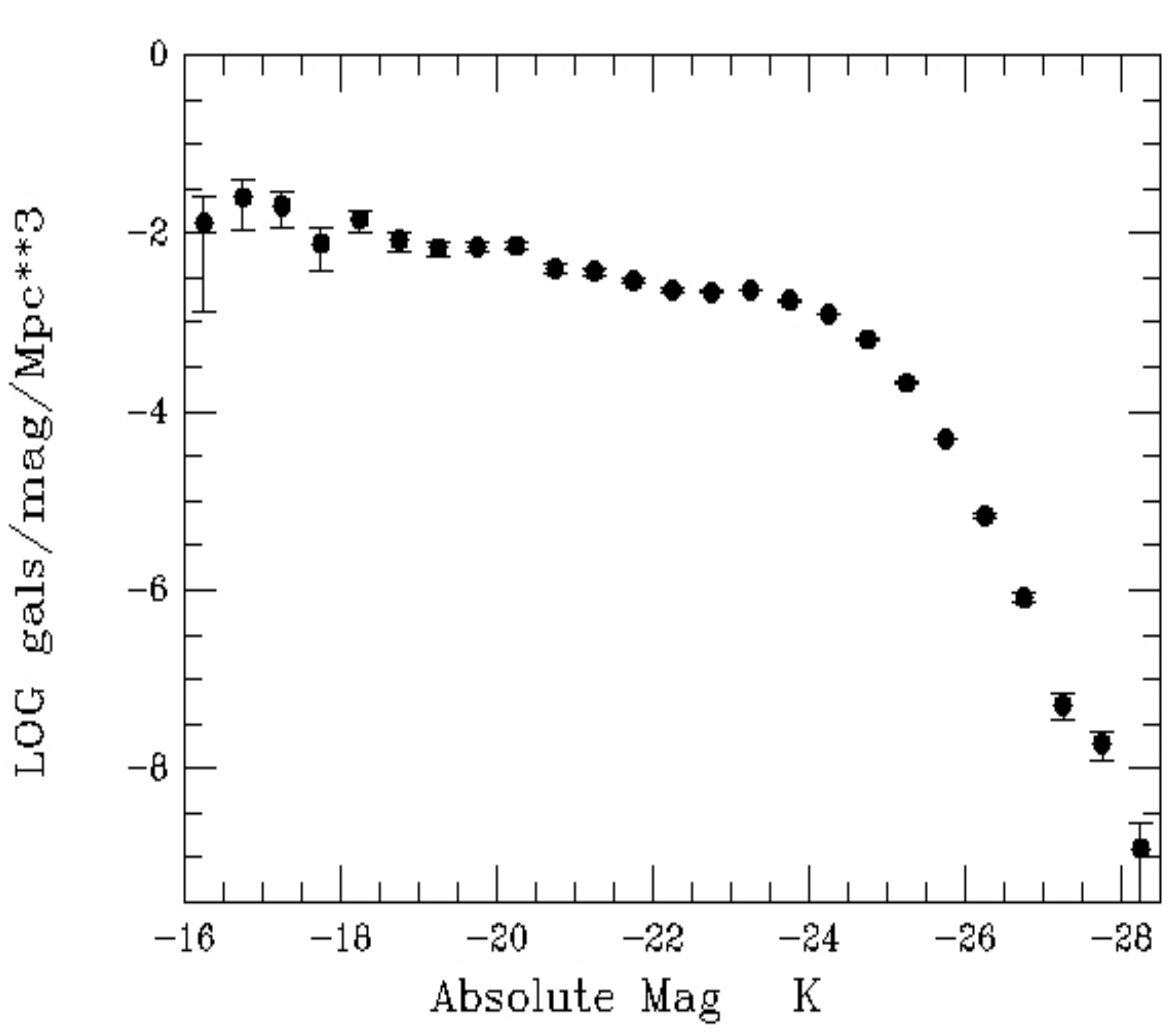

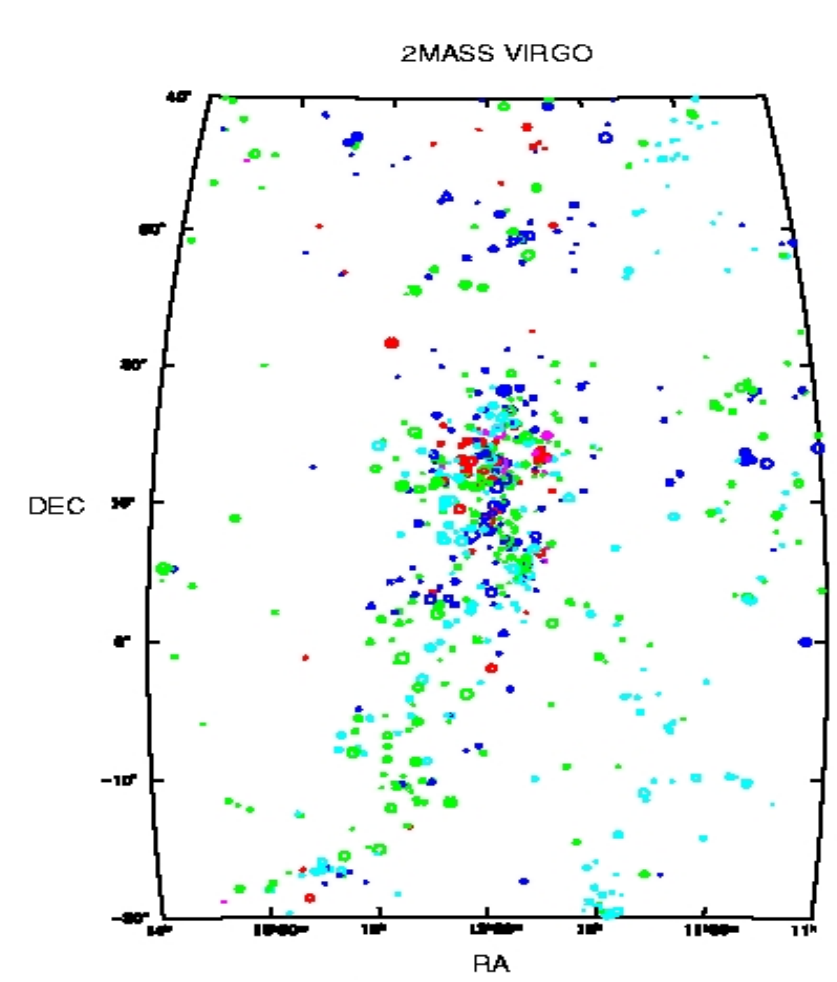

## 9. Galactic Collisions

An upcoming galactic collision is likely to wake the black hole at the centre of the Milky Way and turn our home galaxy into a quasar — it could even fling our solar system into intergalactic space, according to new research.

"This phenomenon will generate powerful jets of high energy radiation emanating from just outside the black hole," lead author Marius Cautun, a researcher at Durham University, said in a press release. "While this will not affect our Solar System, there is a small chance that we might not escape unscathed from the collision between the two galaxies which could knock us out of the Milky Way and into interstellar space."

## 10. Gravitational Lensing

By directly examining the gravitational effects of all the clustering elements in the Universe on light from Background sources,gravitational lensing is a potent technique for Astronomy. Weak lensing, the statistical detection of lensing in the CMB, and the morphologies of background galaxies are the three most important applications in cosmology.

In terms of the power spectrum of the lensing potential, which equates to the weighted integral over the gravitational potential, we calculate the statistics of all lensing observables. The temperature observations of off-diagonal correlations can be used to measure CMB lensing. Similar to polarisation, galaxy form correlations can be broken down. Only E-modes in leading order are sources with lensing. The cross-correlation with foreground galaxies, in addition to the auto-correlations of lensing E- modes is significant for cosmological constraints [14,15].

| Planet | Mass (Jupiter mass) | Radius (Jupiter) | Orbital Period, *P* (days) |
|---|---|---|---|
| OGLE-TR-56 | 1.45 | 1.23 | 1.21 |
| OGLE-TR-113 | 1.35 | 1.08 | 1.43 |
| OGLE-TR-132 | 1.19 | 1.13 | 1.69 |
| OGLE-TR-10 | 0.57 | 1.24 | 3.1 |
| OGLE-TR-111 | 0.53 | 1 | 4.02 |
| TrES-1 | 0.61 | 1.08 | 3.03 |

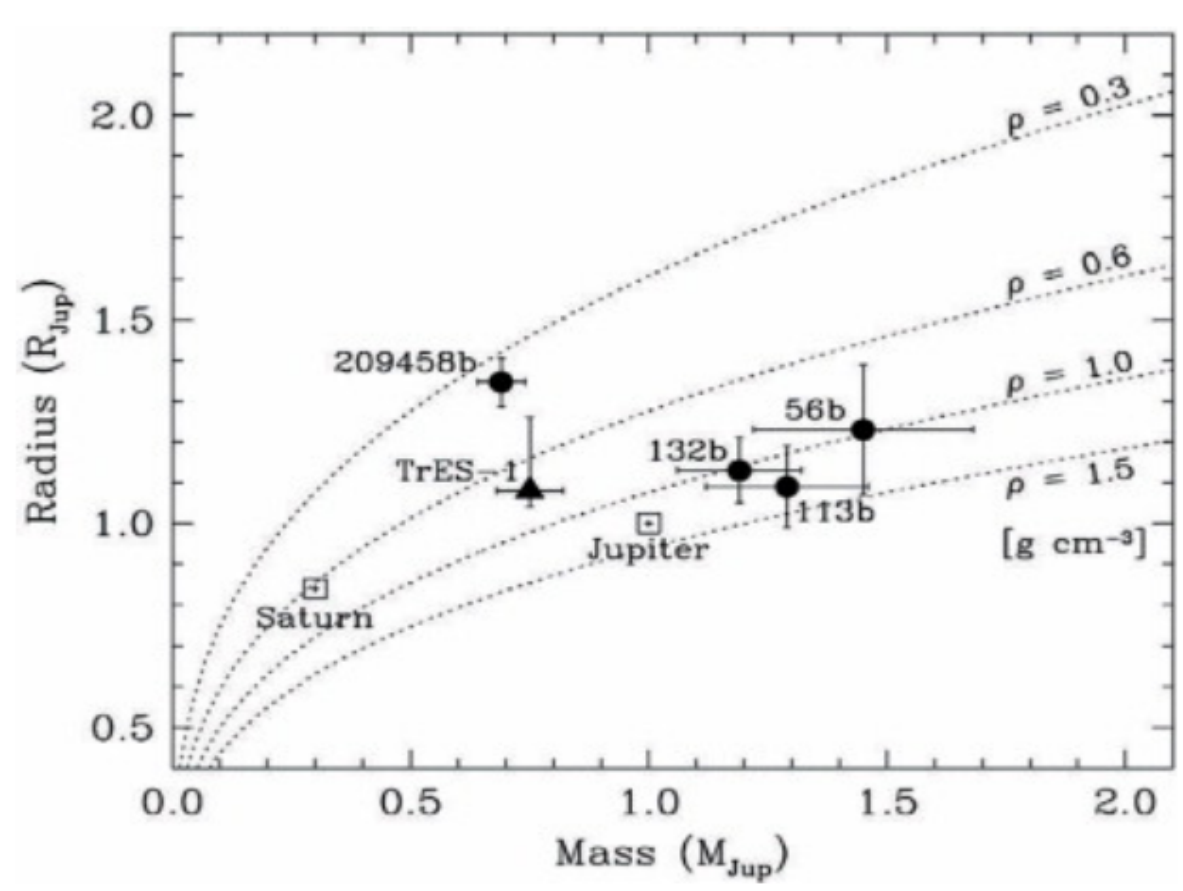

## 11. Weakly Interacting Massive Particles (WIMPs)

WIMPS are heavy, electromagnetically neutral subatomic particles that are thought to make up the majority of dark matter and, consequently, about 22% of the Universe. Because if the dark matter particles were light and fast moving, they would not have clamped together in the density fluctuations from which galaxies and clusters of galaxies formed.

Therefore, it is believed that these particles are heavy and slow moving. These particles are electromagnetically neutral since there is no light coming from them. Weakly interacting large particles is the name given to the particles as a result of these characteristics(WIMPs). Because of the quantity of evidence, WIMPs are “non-baryonic”, or different from baryons (massive particles consisting of three quarks like the protons and neutron). [16]

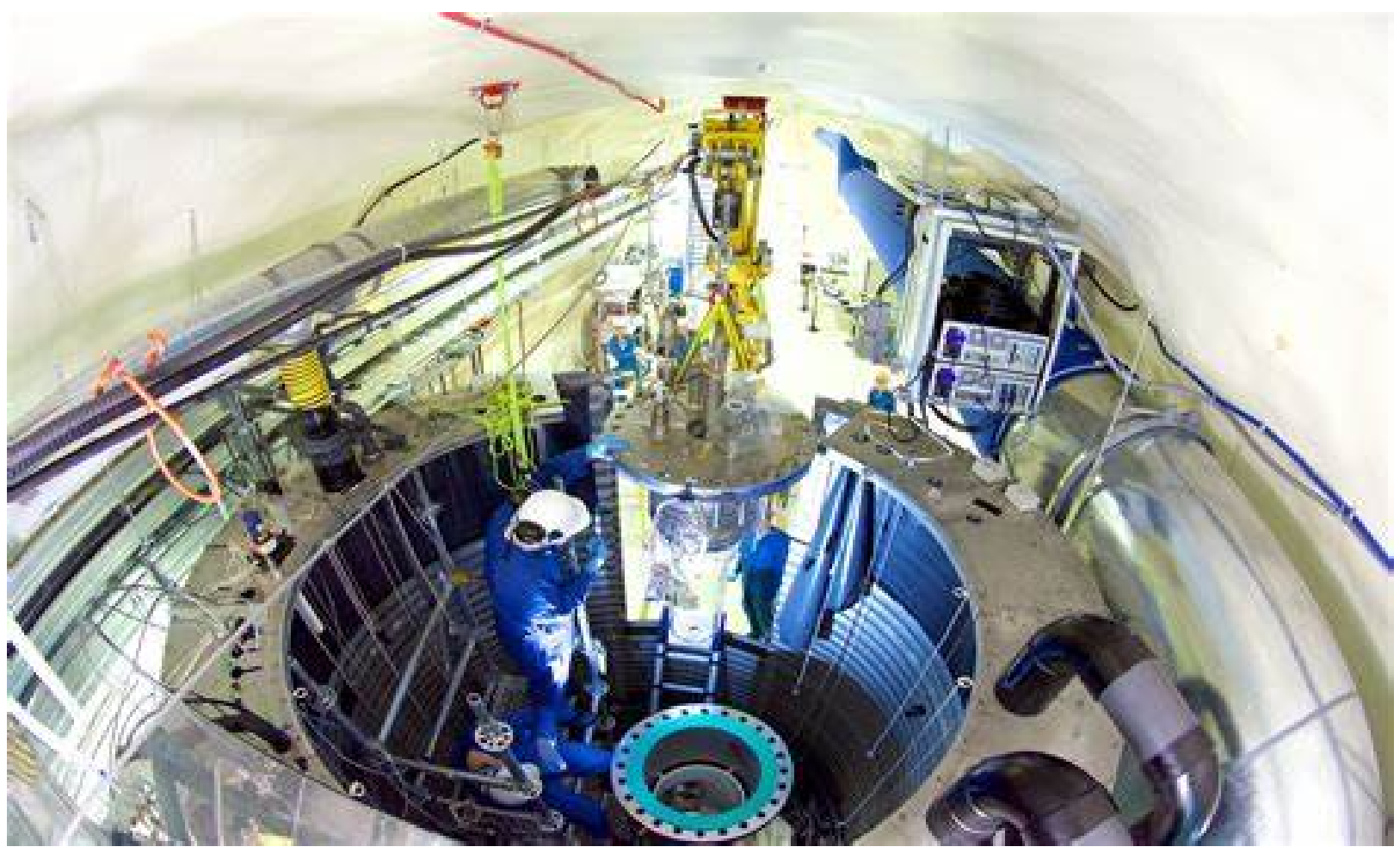

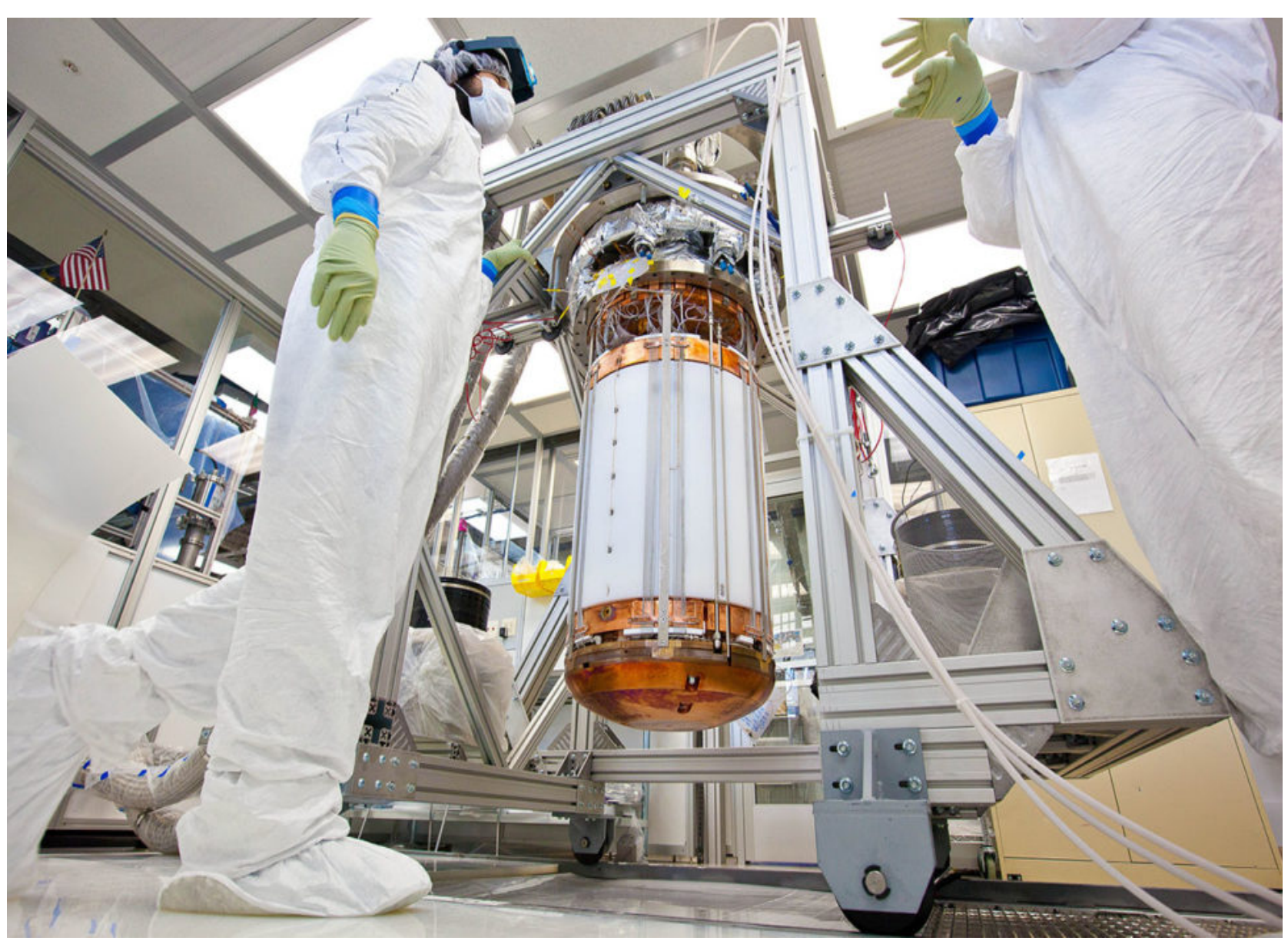

## Conclusion

Although Dark Matter's presence isn't directly observed, it is needed to understand the nature of our universe. Whether it is present or not and whether we need a new way to look at the theory of Gravity, till today it fits many models, simulations and observations too. It has always been there but now it's time that the entity be brought before the world. The accelerating expansion rate of the universe is a robust consequence of dark matter as well as dark energy. Hence, more specific and detailed studies are needed instead of ambiguous & vague ones. Team Astrophysics from Quamacro Research Institute has done their job in presenting the many forms of dark matter and their behaviour.

---

## Acknowledgements

We would like to express our special thanks to Dr. Vishal Kahar for his able guidance and support to help us out in completing this project.

---